\documentclass[prb,aps,twocolumn]{revtex4}

\usepackage {graphicx}
\usepackage {amsmath}
\usepackage {amssymb}

\newcommand* {\Ds}{\displaystyle}
\newcommand* {\Ts}{\textstyle}
\newcommand* {\frack}[2]{{\Ts\frac{#1}{#2}}}

\newcommand* {\ket}[1]{\ensuremath{\left| {#1} \right\rangle}}
\newcommand* {\expect}[1]{\ensuremath{\left\langle {#1} \right\rangle}}
\newcommand* {\ketk}[1]{\ensuremath{| {#1} \rangle}}
\newcommand* {\expectk}[1]{\ensuremath{\langle {#1} \rangle}}
\newcommand* {\matrixelk}[3]{\ensuremath{\langle {#1} | {#2} | {#3} \rangle}}

\usepackage {bm}
\newcommand* {\vek}[1]{\ensuremath{\bm{\mathrm{#1}}}}
\newcommand* {\vekc}[1]{\ensuremath{\bm{\mathcal{#1}}}}

\newcommand* {\kk}{\vek{k}}
\newcommand* {\rr}{\vek{r}}
\newcommand* {\ee}{\ensuremath{\mathrm{e}}}
\newcommand* {\Ee}{\ensuremath{\mathcal{E}}}
\newcommand* {\imag}{\ensuremath{i}}

\begin{document}

\title{Spin Orientation and Spin Precession in Inversion-Asymmetric
Quasi Two-Dimensional Electron Systems}

\author{R.~Winkler}
\affiliation{Institut f\"ur Technische Physik III, Universit\"at
Erlangen-N\"urnberg, Staudtstr. 7, D-91058 Erlangen, Germany}

\date{May 14, 2003}
\begin{abstract}
  Inversion asymmetry induced spin splitting of the electron states
  in quasi two-dimensional (2D) systems can be attributed to an
  effective magnetic field $\vekc{B}$ which varies in magnitude and
  orientation as a function of the in-plane wave vector $\kk_\|$.
  Using a realistic $8\times 8$ Kane model that fully takes into
  account spin splitting because of both bulk inversion asymmetry
  and structure inversion asymmetry we investigate the spin
  orientation and the effective field $\vekc{B}$ for different
  configurations of a quasi 2D electron system.  It is shown that
  these quantities depend sensitively on the crystallographic
  direction in which the quasi 2D system was grown as well as on the
  magnitude and orientation of the in-plane wave vector $\kk_\|$.
  These results are used to discuss how spin-polarized electrons can
  precess in the field $\vekc{B}(\kk_\|)$. As a specific example we
  consider Ga$_{0.47}$In$_{0.53}$As-InP quantum wells.
\end{abstract}
\maketitle

%%%%%%%%%%%%%%%%%%%%%%%%%%%%%%%%%%%%%%%%%%%%%%%%%%%%%%%%%%%%%%%%%%
\section{Introduction}
\label{sec:intro}

Spin degeneracy in a two-dimensional (2D) system is due to the
combined effect of spatial inversion symmetry and time inversion
symmetry. \cite{kit63} If the spatial inversion symmetry is lifted
spin-orbit interaction gives rise to a spin splitting of the
electron states even at a magnetic field $B=0$. In quasi 2D systems
the $B=0$ spin splitting can be caused by the bulk inversion
asymmetry (BIA) of the underlying crystal structure \cite{dre55a} as
well as by the structure inversion asymmetry (SIA) due to, e.g., an
electric field $\Ee$ perpendicular to the plane of the 2D system.
\cite{byc84} The $B=0$ spin splitting is of considerable interest
both because of its importance for our understanding of the
fundamental properties of quasi 2D systems
\cite{dya84,pik84,pik95,pap99} as well as because of possible
applications in the field of spintronics. \cite{wol01}

Common III-V and II-VI semiconductors such as GaAs, InSb, and
HgCdTe, have a zinc blende structure. To lowest order in the wave
vector $\kk$ BIA spin splitting in these systems is characterized by
the so-called Dresselhaus term \cite{dre55a} whereas spin splitting
due to SIA is characterized by the Rashba term. \cite{byc84} Often
the discussion of spin splitting is restricted to these lowest-order
terms. \cite{mal86,pik88,epp88,and92} Spin splitting of higher
orders in $\kk$ can be fully taken into account by the $8 \times 8$
Kane model \cite{tre79} or the $14 \times 14$ extended Kane model.
\cite{ros84} The higher-order terms can be quite important for a
quantitative discussion of $B=0$ spin splitting. \cite{win93a,
wis98}

For a given in-plane wave vector $\kk_\|$ we can always find a spin
axis $\expectk{\vek{S} (\kk_\|)}$ \emph{local in $\kk_\|$ space}
such that we have spin-up and spin-down eigenstates with respect to
the axis $\expectk{\vek{S} (\kk_\|)}$. Note that we cannot call the
spin-split branches $E_\pm (\kk_\|)$ of the energy surface spin-up
or spin-down because the direction of $\expectk{\vek{S}}$ varies as
a function of $\kk_\|$ such that averaged over all occupied states
the branches contain equal contributions of up and down spinor
components. This reflects the fact that in nonmagnetic materials we
have at $B=0$ a vanishing magnetic moment.

The spin orientation $\expectk{\vek{S} (\kk_\|)}$ can be attributed
to an effective magnetic field $\vekc{B} (\kk_\|)$ (Refs.\ 
\onlinecite{dya72,and92}). A discussion of $\expectk{\vek{S}
(\kk_\|)}$ based on the lowest-order terms in the effective
spin-orbit interaction has previously been given by several authors,
see, e.g., Refs.\ \onlinecite{dat90,sch98a,ham00,mat02,sch03}. In
the present paper we compare these results with our calculations of
$\expectk{\vek{S} (\kk_\|)}$ and the field $\vekc{B} (\kk_\|)$ using
the more realistic $8 \times 8$ Kane model \cite{tre79} that
takes into account both SIA and BIA up to all orders in $\kk_\|$. It
will be shown that for larger $\kk_\|$ the higher-order terms
result in important modifications of $\expectk{\vek{S} (\kk_\|)}$
and $\vekc{B} (\kk_\|)$.

Datta and Das have proposed a novel spin transistor \cite{dat90}
where the current modulation arises from the precession of
spin-polarized electrons in the effective field $\vekc{B} (\kk_\|)$,
while ferromagnetic contacts are used to preferentially inject and
detect specific spin orientations. Recently, extensive research
aiming at the realization of such a device has been under way.
\cite{pasps01} Here we will use our results for the field $\vekc{B}
(\kk_\|)$ in order to discuss spin precession and its tunability for
different device configurations. It will be shown that for certain
configurations the precession of spin-polarized electrons is
determined only by the tunable SIA spin splitting; but it is
essentially independent of the magnitude of BIA spin splitting. For
other configurations the tunability of spin precession is
significantly suppressed due to the interplay of SIA and~BIA.

We would like to emphasize that the present results apply only to
electrons with an (effective) spin $j=1/2$. Holes in the topmost
valence band, on the other hand, have an effective spin $j=3/2$
(Ref.\ \onlinecite{lut56}). Therefore, spin orientation and spin
precession in quasi 2D hole systems is qualitatively different from
spin orientation and spin precession in quasi 2D electron systems.
Hole systems will thus be covered in a future publication.

%%%%%%%%%%%%%%%%%%%%%%%%%%%%%%%%%%%%%%%%%%%%%%%%%%%%%%%%%%%%%%%%%
\section{Spin Orientation of 2D Electron States}
\label{sec:spinpol}

In the following we want to discuss the wave vector dependent spin
orientation $\expectk{\vek{S} (\kk_\|)}$ for different models of
spin splitting. We will compare the analytical results for the
Rashba model and Dresselhaus model with our more realistic
calculations based on the $8 \times 8$ Kane Hamiltonian that takes
into account SIA and BIA spin splitting up to all orders in
$\kk_\|$.

%%%%%%%%%%%%%%%%%%%%%%%%%%%%%%%%%%%%%%%%%%%%%%%%%%%%%%%%%%%%%%%%%%
\subsection{General Discussion}
\label{sec:spinpol_general}

\begin{figure}[tbp]
  \includegraphics[width=\columnwidth]{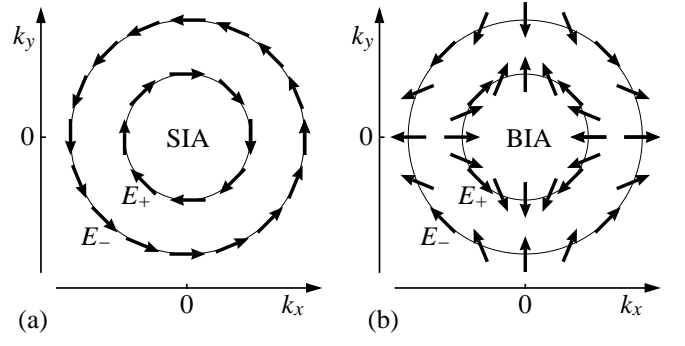}
  \caption[]{\label{fig:sbia_pol} Lowest order spin orientation
  $\expectk{\vek{\sigma}}$ of the eigenstates $\ketk{\psi_\pm
  (\kk_\|)}$ in the presence of (a) SIA and (b) BIA. The inner
  (outer) circle shows $\expectk{\vek{\sigma}}$ along contours of
  constant energy for the upper (lower) branch $E_+$ ($E_-$) of the
  spin-split dispersion.}
\end{figure}

First we want to discuss the spin orientation in the presence of
SIA. Here to lowest order in the in-plane wave vector $\kk_\| =
(k_x, k_y,0)$ the spin splitting is characterized by the Rashba
Hamiltonian~\cite{byc84}
\begin{equation}
\label{rashba}
H_\mathrm{SIA} =  \alpha \left(\sigma_x k_y - \sigma_y k_x
  \right) ,
\end{equation}
where $\sigma_x$ and $\sigma_y$ are Pauli spin matrices and $\alpha$
is a prefactor that depends on the constituting materials and on the
geometry of the quasi 2D system. If we use polar coordinates for the
in-plane wave vector, $\kk_\| = k_\| (\cos \varphi, \sin\varphi,
0)$, the spin splitting is given by
\begin{equation}
\label{eq:rash_disp}
E^\mathrm{SIA}_\pm (\kk_\|) = \pm \alpha k_\|
\end{equation}
independent of the angle $\varphi$ and the eigenstates are
\begin{equation}
  \label{eq:rashba_eigen}
  \ketk{\psi^\mathrm{SIA}_\pm (\kk_\|)} =
  \frac{\ee^{\imag \kk_\| \rr_\|}}{2 \pi} \;  \xi_{\kk_\|} (z) \;
  \frac{1}{\sqrt{2}}
    \ket{\begin{array}{c} 1 \\ \mp \imag \ee^{\imag\varphi} \end{array}}
\end{equation}
with $\rr_\| = (x,y,0)$ and envelope functions $\xi_{\kk_\|} (z)$.
In Eq.\ (\ref{eq:rashba_eigen}) we have assumed that the Rashba
coefficient $\alpha$ is positive. The spin orientation of the
eigenstates (\ref{eq:rashba_eigen}) is given by the expectation
value with respect to the vector $\vek{\sigma}$ of Pauli spin
matrices.
\begin{subequations}
  \label{eq:spin_rashba}
\begin{eqnarray}
  \expect{\vek{\sigma} (\kk_\|)}_\pm 
  & \equiv & \big\langle\psi_\pm (\kk_\|) \,
  \big|\,\rule{0pt}{2.5ex} \vek{\sigma}\, \big| \,
  \psi_\pm (\kk_\|) \big\rangle \\[1ex]
  & = & \left(\begin{array}{c}
      \pm \sin\varphi \\[1ex] \mp \cos\varphi \\[1ex] 0
    \end{array} \right)
  = \pm \left(\begin{array}{c}
      \cos\left(\varphi - {\Ts\frac{\pi}{2}}\right) \\[1ex]
      \sin\left(\varphi - {\Ts\frac{\pi}{2}}\right) \\[1ex]
      0
    \end{array} \right).
\end{eqnarray}
\end{subequations}
Note that Eq.\ (\ref{eq:spin_rashba}) is independent of
the envelope function $\xi_{\kk_\|} (z)$ and the magnitude $k_\|$ of
the in-plane wave vector. The spin orientation
(\ref{eq:spin_rashba}) of the eigenfunctions (\ref{eq:rashba_eigen})
as a function of the direction of the in-plane wave vector is
indicated by arrows in Fig.~\ref{fig:sbia_pol}(a).

Next we want to discuss the spin orientation in the presence of BIA
spin splitting. For quasi 2D systems in a quantum well (QW) grown in
the crystallographic direction [001] the Dresselhaus term becomes
\cite{mal86, epp88}
\begin{equation}
  \label{eq:bia_2d}
  H_\mathrm{BIA} =
  \eta \left[\sigma_x k_x (k_y^2 - \expectk{k_z^2})
           + \sigma_y k_y (\expectk{k_z^2} - k_x^2)
    \right]
\end{equation}
with a material-specific coefficient $\eta$. This equation can
easily be diagonalized. We obtain a spin splitting
\begin{subequations}
  \label{eq:e_bia_2d}
\begin{eqnarray}
  E^\mathrm{BIA}_\pm (\kk_\|) & = & \pm \eta \, k_\|
  \sqrt{\expectk{k_z^2}^2 + \big( \frack{1}{4} k_\|^2 - \expectk{k_z^2}\big)
        k_\|^2 \sin (2\varphi)^2}
  \nonumber\\ \\
  % \\[2.8ex]
  & \approx & \pm \eta
  \, \expectk{k_z^2} \, k_\| \pm \mathcal{O} (\kk_\|^3).
  \label{eq:e_bia_2d_tay}
\end{eqnarray}  
\end{subequations}
We see here that in leading order of $\kk_\|$ the Dresselhaus term
(\ref{eq:bia_2d}) gives rise to a spin splitting independent of the
direction of $\kk_\|$ that is apparently very similar to the Rashba
spin splitting (\ref{eq:rash_disp}). Nevertheless, the corresponding
wave functions are qualitatively different due to the different
symmetries of the terms (\ref{rashba}) and (\ref{eq:bia_2d}).  If we
neglect the terms cubic in $\kk_\|$ the eigenfunctions in the
presence of Dresselhaus spin splitting are
\begin{equation}
  \label{eq:dressel_eigen}
  \ketk{\psi^\mathrm{BIA}_\pm (\kk_\|)} =
  \frac{\ee^{\imag \kk_\| \rr_\|}}{2 \pi} \;  \xi_{\kk_\|} (z) \;
  \frac{1}{\sqrt{2}} \ket{\begin{array}{c}
    1 \\ \mp \ee^{- \imag\varphi} 
  \end{array}}
\end{equation}
so that
\begin{equation}
  \label{eq:spin_dressel}
  \expect{\vek{\sigma} (\kk_\|)}_\pm 
  = \mp \left(\begin{array}{c}
      \cos\left(- \varphi\right) \\
      \sin\left(- \varphi\right) \\
      0
    \end{array} \right).
\end{equation}
The spin orientation (\ref{eq:spin_dressel}) of the eigenfunctions
(\ref{eq:dressel_eigen}) as a function of the direction of the
in-plane wave vector is indicated by arrows in
Fig.~\ref{fig:sbia_pol}(b). For the Rashba spin splitting we see in
Fig.~\ref{fig:sbia_pol}(a) that if we are moving clockwise on a
contour of constant energy $E (\kk_\|)$ the spin vector is rotating
in the same direction, consistent with the axial symmetry of the
Rashba term. On the other hand, Eq.\ (\ref{eq:spin_dressel}) and
Fig.~\ref{fig:sbia_pol}(b) show that in the presence of BIA the spin
vector is rotating counterclockwise for a clockwise motion in
$\kk_\|$ space.

In the above discussion we have assumed that the wave functions are
two-component spinors. In general, the quasi 2D eigenstates of a
multiband Hamiltonian are of the form \cite{alt85a}
\begin{equation}
\label{eq:envelope_qw}
\ketk{\psi (\kk_\|)} =
\frac{\ee^{\imag \kk_\| \rr_\|}}{2 \pi} \;
\sum_{n} \xi_{n\kk_\|} (z) \;
u_{n} (\rr)
\end{equation}
with envelope functions $\xi_{n\kk_\|} (z)$, and $u_{n} (\rr)$
denotes the band edge Bloch function of the $n$th bulk band. Here we
must evaluate the expectation value of
\begin{equation}
 \vek{S} = \vek{\sigma} \otimes \openone_\mathrm{orb} ,
\end{equation}
where the identity operator~$\openone_\mathrm{orb}$ refers to the
orbital part of $\ketk{\psi (\kk_\|)}$. For the $8 \times 8$ Kane
model \cite{tre79} containing the bands $\Gamma_6^c$, $\Gamma_8^v$,
and $\Gamma_7^v$ we obtain for $i=x,y,z$
\begin{equation}
  \label{eq:spin_op_nondiag}
  S_i = \left(
    \renewcommand{\arraystretch}{1.2}
    \begin{array}{ccc}
      \sigma_i & 0 & 0 \\
      0 & \frack{2}{3} J_i & -2 U_i \\
      0 & -2 U_i^\dagger &  -\frack{1}{3}\sigma_i
    \end{array} \right),
\end{equation}
where $J_i$ denotes the matrices for angular momentum $j=3/2$, and
the matrices $U_i$ are defined in Ref.~\onlinecite{tre79}. Once
again the expectation value $\matrixelk{\psi}{\vek{S}}{\psi}$ is a
three-component vector that can be identified with the spin
orientation of the multicomponent wave function $\ketk{\psi}$. We
remark that while the vector $\expectk{\vek{\sigma}}$ of a spin
$1/2$ system is always strictly normalized to unity, this condition
is in general not fulfilled for the spin expectation value
$\expectk{\vek{S}}$ of multicomponent single particle states. This
is due to the fact that in the presence of spin-orbit interaction we
cannot factorize the multicomponent wave function
(\ref{eq:envelope_qw}) into an orbital part and a spin part.
However, for electrons the deviation of $|\expectk{\vek{S}}|$ from
unity is rather small (typically less than 1\%) so that it is
neglected here.

For free electrons in the presence of an external magnetic field
$\vek{B}$ the unit vector $\expectk{\vek{\sigma}}$ is parallel to
the vector $\vek{B}$. Following this picture we can attribute the
$B=0$ spin splitting in quasi 2D systems to an effective magnetic
field $\vekc{B} (\kk_\|)$ parallel to $\expectk{\vek{S} (\kk_\|)}$.
Obviously the magnitude of this effective magnetic field should be
related to the magnitude of the $B=0$ spin splitting. However,
depending on the particular problem of interest it can be convenient
to define the magnitude of spin splitting in two different ways: The
energy difference $\Delta E = E_+ (\kk_\|) - E_- (\kk_\|)$
characterizes the magnitude of spin splitting for a given wave
vector $\kk_\|$ whereas the wave vector difference $\Delta k$
characterizes the magnitude of spin splitting at a fixed energy $E$.
While the former is relevant, e.g., for Raman experiments,
\cite{wis98} the latter quantity is an important parameter, e.g.,
for spin relaxation \cite{pik84,dya72} and for the spin transistor
proposed by Datta and Das. \cite{dat90}

\begin{figure}[tbp]
  \includegraphics[width=0.60\columnwidth]{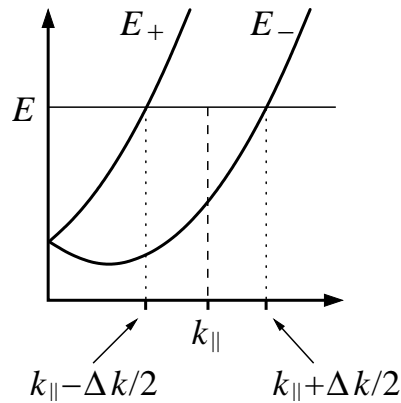}
  \caption[]{\label{fig:rashba_edisp_pmdk} For the given
  energy $E$ and a fixed direction of the in-plane wave vector
  $\kk_\|$ we determine $\kk_\| \mp \Delta \kk / 2$ such that $E =
  E_+ (\kk_\| - \Delta \kk / 2) = E_- (\kk_\| + \Delta \kk / 2)$.
  Here $E_+$ ($E_-$) denotes the upper (lower) branch of the
  spin-split dispersion.}
\end{figure}

In the following we want to explore the second definition where the
effective magnetic field is given by $\vekc{B} =
\expectk{\vek{S}} \Delta k$. Our precise definition of $\Delta k$ is
illustrated in Fig.~\ref{fig:rashba_edisp_pmdk}: For the given
energy $E$ and a fixed direction $\varphi$ of the in-plane wave
vector $\kk_\| = k_\| (\cos \varphi, \sin\varphi, 0)$ we determine
$\kk_\| \mp \Delta \kk / 2$ such that $E = E_+ (\kk_\| - \Delta \kk /
2) = E_- (\kk_\| + \Delta \kk / 2)$. Here $E_+$ ($E_-$) denotes the
upper (lower) branch of the spin-split dispersion. Then we define
\begin{equation}
  \label{eq:b_eff_def}
  \vekc{B}
  =   \expectk{\vek{S}}_+ \,\Delta k
  = - \expectk{\vek{S}}_- \,\Delta k
\end{equation}
with the sign convention that the field $\vekc{B}$ is
parallel to the effective field felt by the electrons in the upper
branch $E_+ (\kk_\|)$ and we have used the short-hand notation
\begin{equation}
  \label{eq:S_exp_abbrev}
  \expectk{\vek{S}}_\pm =
  \big\langle\psi_\pm (\kk_\| \mp \Delta \kk/2) \,
  \big|\,\rule{0pt}{2.5ex} \vek{S}\, \big| \,
  \psi_\pm (\kk_\| \mp \Delta \kk/2) \big\rangle.
\end{equation}
We remark that for a parabolic band with effective mass $m^\ast$
plus Rashba term (\ref{rashba}) the wave vector difference $\Delta
k$ can be evaluated analytically \cite{dat90}
\begin{equation}
  \label{eq:rashba_dk}
  \Delta k_\mathrm{Rashba} = \frac{2 m^\ast \, \alpha}{\hbar^2}
\end{equation}
independent of the magnitude of $k_\|$. From an experimental point
of view it should be kept in mind that spin splitting is often
measured by analyzing Shubnikov--de Haas oscillations, see, e.g.,
Refs.\ \onlinecite{luo88, das89, eng97, gru00}. Such experiments
yield spin subband densities $N_\pm$ which are directly related
to~$\Delta k$
\begin{equation}
  \label{eq:dk_exp}
  \Delta k = \sqrt{4\pi} \,\big(\sqrt{N_-} - \sqrt{N_+} \big),
\end{equation}
provided we can ignore anisotropic contributions to \mbox{$B=0$}
spin splitting. (However, see also Refs.~\onlinecite{win00,kep02}.)

The definition (\ref{eq:b_eff_def}) presupposes that the spin
expectation values $\expectk{\vek{S}}_+$ and $\expectk{\vek{S}}_-$
are strictly antiparallel to each other. In Eq.\ 
(\ref{eq:spin_rashba}) we saw that for the Rashba Hamiltonian this
condition is fulfilled exactly. This is closely related to the fact
that for the Rashba Hamiltonian the spin subband eigenstates
$\ketk{\psi^\mathrm{SIA}_+ (\kk_\|)}$ and $\ketk{\psi^\mathrm{SIA}_-
(\kk'_\|)}$ are orthogonal -- independent of the magnitude of
$\kk_\|$ and $\kk'_\|$ as long as the wave vectors $\kk_\|$ and
$\kk'_\|$ are parallel to each other. \cite{irep} In general,
$\ketk{\psi_+ (\kk_\| - \Delta \kk / 2)}$ and $\ketk{\psi_- (\kk_\|
+ \Delta \kk / 2)}$ are only approximately orthogonal so that
$\expectk{\vek{S}}_+$ and $\expectk{\vek{S}}_-$ are only
approximately antiparallel. However, we find that the angle between
the vectors $\expectk{\vek{S}}_+$ and~$\expectk{\vek{S}}_-$ is
always very close to 180$^\circ$ with an error $\lesssim 1^\circ$ so
that we neglect this point in the remaining discussion.

Even though we can evaluate the spin expectation value
$\expectk{\vek{S}}$ for each spin subband separately we do not
attempt to define an effective magnetic field $\vekc{B}$ for each
spin subband. This is due to the fact that $\vekc{B}$ is commonly
used to discuss phenomena like spin relaxation \cite{pik84,dya72}
and spin precession \cite{dat90} (see below) which cannot be
analyzed for each spin subband individually.

The allowed directions of the effective magnetic field $\vekc{B}$
can readily be deduced from the symmetry of the QW. The spin-split
states for a fixed wave vector $\kk_\|$ are orthogonal to each
other, i.e., the spin vectors of these states are antiparallel. The
spin orientation of eigenstates for different wave vectors in the
star of $\kk_\|$ are connected by the symmetry operations of the
system. \cite{bir74} Accordingly, only those spin orientations of
the spin-split eigenstates are permissible for which every symmetry
operation maps orthogonal states onto orthogonal states. In a QW
grown in the crystallographic direction [001] the effective field
$\vekc{B}$ is parallel to the plane of the quasi 2D system. Indeed,
the field $\vekc{B}$ due to SIA is always in the plane of the well.
For growth directions other than [001], the effective field due to
BIA has, however, also an out-of-plane component. In particular, a
symmetric QW grown in the crystallographic direction [110] has the
point group $C_{2v}$. \cite{c2v} Here the BIA induced field
$\vekc{B} (\kk_\|)$ must be perpendicular to the plane of the QW (to
all orders in $\kk_\|$). This situation is remarkable because
D'yakonov-Perel' spin relaxation is suppressed if the spins are
oriented perpendicular to the 2D plane. \cite{dya86,ohn99} Note also
that in [110] grown QW's $\vekc{B}$ vanishes for $\kk_\| \parallel
[001]$ because here the group of $\kk_\|$ is $C_{2v}$ which has
merely one irreducible double group representation, $\Gamma_5$,
which is two-dimensional. \cite{par55}

%%%%%%%%%%%%%%%%%%%%%%%%%%%%%%%%%%%%%%%%%%%%%%%%%%%%%%%%%%%%%%%%%%
\subsection{Numerical Results}
\label{sec:spinpol_numeric}

\begin{figure*}[tbp]
  \includegraphics[width=\textwidth]{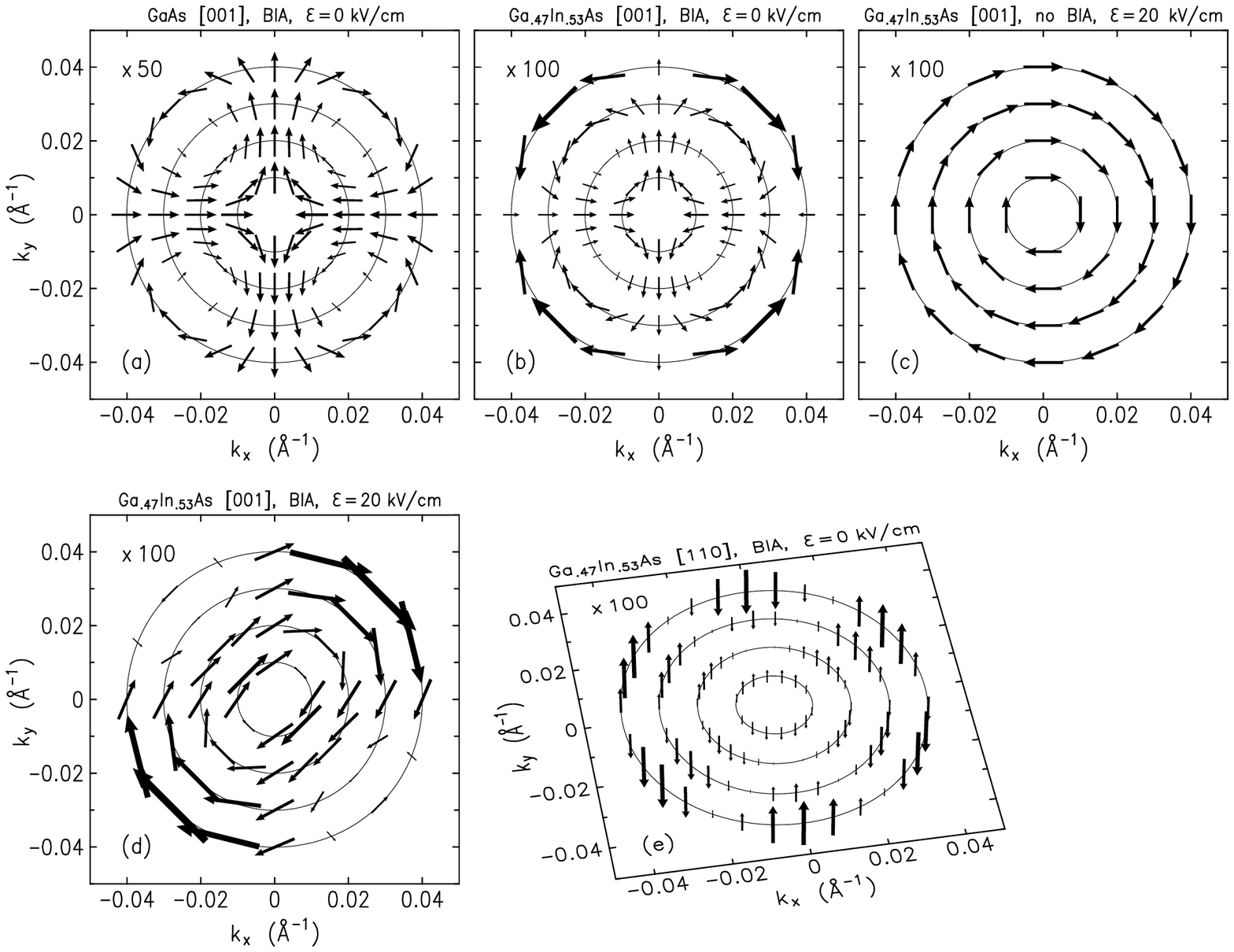}

  \caption[]{\label{fig:beff_gainas} Effective magnetic field
  $\vekc{B} (\kk_\|)$ for (a) a GaAs-Al$_{0.3}$Ga$_{0.7}$As QW and
  (b-e) a Ga$_{0.47}$In$_{0.53}$As-InP QW, both with a well width of
  100~{\AA}. In (a), (b), and (e) we assume that we have a symmetric
  well with BIA spin splitting only.  (c) shows $\vekc{B} (\kk_\|)$
  due to an external field of $\Ee = 20$~kV/cm but neglecting BIA
  while (d) shows $\vekc{B} (\kk_\|)$ when we have both BIA and SIA
  spin splitting (again for $\Ee = 20$~kV/cm). While (a-d) refers to
  a QW grown in the crystallographic direction [001] we have assumed
  in (e) that the QW was grown in [110] direction. The dimensions of
  the arrows are proportional to $|\vekc{B}| = \Delta k$. For
  Ga$_{0.47}$In$_{0.53}$As, we have amplified $\vekc{B} (\kk_\|)$ by
  a factor of 100, for GaAs it has been scaled by a factor of~50.
  All calculations are based on an $8\times 8$ Kane Hamiltonian
  ($\Gamma_6^c$, $\Gamma_8^v$, and $\Gamma_7^v$) including
  off-diagonal remote band contributions of second order in~$\kk$
  (Ref.~\onlinecite{tre79,win93a}).}
\end{figure*}

The analytically solvable models (\ref{rashba}) and
(\ref{eq:bia_2d}) allow one to study the qualitative trends of BIA
and SIA spin splitting in quasi 2D systems. The largest spin
splitting can be achieved in narrow-gap semiconductors where the
subband dispersion is highly nonparabolic. Therefore, we present
next numerically calculated results for $\vekc{B} (\kk_\|)$ obtained
by means of an accurate $8\times 8$ Kane Hamiltonian ($\Gamma_6^c$,
$\Gamma_8^v$, and $\Gamma_7^v$) including off-diagonal remote band
contributions of second order in~$\kk$
(Refs.~\onlinecite{tre79,win93a}). First we analyze BIA spin
splitting that is always present in zinc blende QW's. In
Fig.~\ref{fig:beff_gainas}(a) we show the effective field
(\ref{eq:b_eff_def}) along contours of constant energy for a
symmetric GaAs QW grown in the crystallographic direction $[001]$
with a well width of 100~{\AA}. The dimensions of the arrows in
Fig.~\ref{fig:beff_gainas} are proportional to $|\vekc{B}| = \Delta
k$. We remark that typical Fermi wave vectors of quasi 2D systems
are of the order of the in-plane wave vectors covered in
Fig.~\ref{fig:beff_gainas}.

For small in-plane wave vectors $\kk_\|$ the effective field in
Fig.~\ref{fig:beff_gainas}(a) is well described by Eq.\ 
(\ref{eq:spin_dressel}). For larger wave vectors the effective field
becomes strongly dependent on the direction of $\kk_\|$. In
particular, we see that for $\kk_\| \parallel [110]$ the effective
field reverses its direction when we increase $k_\|$. This reversal
reflects the breakdown of the linear approximation in Eq.\ 
(\ref{eq:e_bia_2d}). For wider wells this breakdown occurs at even
smaller wave vectors $k_\|$, consistent with Eq.\ 
(\ref{eq:e_bia_2d}). 

More specifically, Eq.\ (\ref{eq:e_bia_2d}) predicts for $\kk_\|
\parallel [110]$ a reversal of the direction of $\vekc{B} (\kk_\|)$
when $k_\|^2 = 2\expectk{k_z^2}$, independent of the material
specific coefficient $\eta$. Note, however, that $\expectk{k_z^2}$
depends on the material specific band offset at the interfaces. For
the system in Fig.~\ref{fig:beff_gainas}(a) we find in good
agreement with Eq.\ (\ref{eq:e_bia_2d}) that the reversal of
$\vekc{B} (\kk_\|)$ occurs for $k_\| \approx \sqrt{2\expectk{k_z^2}}
\approx 0.029$~{\AA}$^{-1}$. For comparison, we show in
Fig.~\ref{fig:beff_gainas}(b) the effective field $\vekc{B}
(\kk_\|)$ for a symmetric Ga$_{0.47}$In$_{0.53}$As QW with the same
well width 100~{\AA} like in Fig.~\ref{fig:beff_gainas}(a). Even
though BIA spin splitting is smaller in Ga$_{0.47}$In$_{0.53}$As
than in GaAs, higher-order corrections are more important in
Ga$_{0.47}$In$_{0.53}$As due to the smaller fundamental gap of this
material. Here we have $\expectk{k_z^2} \approx 3.6 \times
10^{-4}$~{\AA}$^{-2}$ so that $\sqrt{2\expectk{k_z^2}} \approx
0.027$~{\AA}$^{-1}$.  On the other hand, the reversal of the
direction of $\vekc{B} (\kk_\|)$ occurs for $k_\| \approx
0.021$~{\AA}$^{-1}$. This illustrates the effect of higher orders in
BIA spin splitting that were neglected in Eq.\ (\ref{eq:bia_2d}) but
fully taken into account in the numerical calculations in
Fig.~\ref{fig:beff_gainas}. [Note that in
Fig.~\ref{fig:beff_gainas}(a) the effective field $\vekc{B}$ has
been amplified by a factor of 50 whereas in
Fig.~\ref{fig:beff_gainas}(b) it has been amplified by a factor
of~100.]

Ga$_{0.47}$In$_{0.53}$As QW's can have a significant Rashba spin
splitting \cite{nit97} so that these systems are of interest for
realizing the spin transistor proposed by Datta and Das.
\cite{dat90} In Fig.~\ref{fig:beff_gainas}(c) we show the effective
field $\vekc{B} (\kk_\|)$ for the same well like in
Fig.~\ref{fig:beff_gainas}(b) assuming that we have SIA spin
splitting due to an electric field $\Ee = 20$~kV/cm, but all
tetrahedral terms that give rise to BIA spin splitting were
neglected. The numerical results are in good agreement with what one
expects according to Eqs.\ (\ref{eq:spin_rashba}) and
(\ref{eq:rashba_dk}). Figure~\ref{fig:beff_gainas}(d) shows the
effective field $\vekc{B} (\kk_\|)$ for a Ga$_{0.47}$In$_{0.53}$As
QW when we have both BIA and SIA spin splitting. Due to the
vectorial character of $\vekc{B}$ we have regions in $\kk_\|$ space
where the contributions of BIA and SIA are additive whereas in other
regions the spin splitting decreases due to the interplay of BIA and
SIA. This is consistent with the well-known fact that in the
presence of both BIA and SIA the spin splitting is anisotropic even
in the lowest order of $\kk_\|$ (Ref.\ \onlinecite{and92}). Using
Eqs.\ (\ref{rashba}) and (\ref{eq:bia_2d}) we obtain
\begin{widetext}
  \begin{subequations}
\label{eq:bia_sia}
\begin{eqnarray}
      E^\mathrm{BIA+SIA}_\pm & = &
      {} \pm k_\| \sqrt{
    \Ds \alpha^2 
    + \alpha \eta
        \big(k_\|^2 - 2 \expectk{k_z^2}\big)
        \sin (2\varphi)
    \Ds + \eta^2 \Big[\expectk{k_z^2}^2 + 
        \big(\frack{1}{4} k_\|^2 - \expectk{k_z^2}\big)
             k_\|^2 \sin (2\varphi)^2
        \Big] }
    \\%[1ex]
    & \approx &
     {} \pm k_\| \sqrt{\alpha^2
    - 2 \alpha \eta
        \expectk{k_z^2} \sin (2\varphi)
    +   \eta^2 \expectk{k_z^2}^2}
         \pm \mathcal{O} (\kk_\|^3).
\end{eqnarray}    
  \end{subequations}
\end{widetext}

In Figs.~\ref{fig:beff_gainas}(a-d) we have considered QW's grown in
the crystallographic direction [001] so that the effective field
$\vekc{B} (\kk_\|)$ is always in the plane of the QW. For
comparison, we show in Fig.~\ref{fig:beff_gainas}(e) the effective
field $\vekc{B} (\kk_\|)$ for a symmetric Ga$_{0.47}$In$_{0.53}$As
QW grown in the crystallographic direction [110] with $k_x \parallel
[00\overline{1}]$ and $k_y \parallel [\overline{1}10]$. Here
$\vekc{B} (\kk_\|)$ is perpendicular to the plane of the QW.
\cite{dya86} For asymmetric QW's grown in the crystallographic
direction [110] the effective field $\vekc{B} (\kk_\|)$ is given by
a superposition of an in-plane field as in
Fig.~\ref{fig:beff_gainas}(c) and a perpendicular field as in
Fig.~\ref{fig:beff_gainas}(e).

%%%%%%%%%%%%%%%%%%%%%%%%%%%%%%%%%%%%%%%%%%%%%%%%%%%%%%%%%%%%%%%%%%
\section{Spin Precession of 2D Electron States}

\subsection{Datta Spin Transistor}

\begin{figure}[tbp]
  \includegraphics[width=0.9\columnwidth]{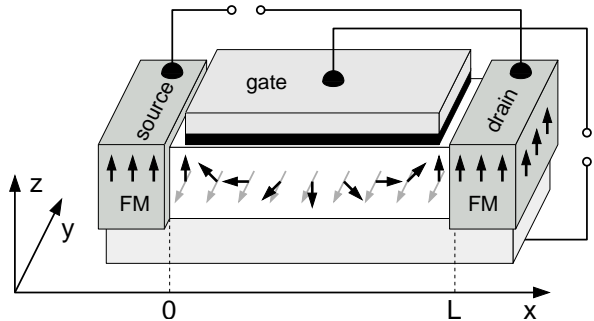}
  \caption[]{\label{fig:datrans} Qualitative sketch of a Datta spin
  transistor. \cite{dat90} Black arrows indicate the spin
  polarization in the ferromagnetic contacts (FM) and the
  semiconducting channel (white). Gray arrows indicate the effective
  magnetic field $\vekc{B} (k_x)$ in the semiconducting channel. A
  top gate is used to tune the spin precession by applying an
  electric field $\Ee$ perpendicular to the semiconducting channel.}
\end{figure}

We want to briefly recapitulate the mode of operation of the spin
transistor proposed by Datta and Das \cite{dat90} (see
Fig.~\ref{fig:datrans}). We assume that the semiconducting channel
between the ferromagnetic contacts is pointing in $x$ direction,
i.e., electrons travel with a wave vector $\kk_\| = (k_x,0,0)$ from
source to drain. A gate in $z$ direction gives rise to a tunable
Rashba coefficient $\alpha$. In this subsection we want to ignore
the Dresselhaus spin splitting (\ref{eq:bia_2d}). When the
spin-polarized electrons in the ferromagnetic source contact are
injected at $x=0$ into the semiconducting channel we must expand its
wave function $\ket{\psi_i}$ in terms of the spin-split eigenstates
$\ketk{\psi^\mathrm{SIA}_\pm (k_x)}$. Here it is the basic idea of
the spin transistor that the polarization of the electrons in the
source contact is chosen perpendicular to $\vekc{B} (k_x) =
(0,\mathcal{B}_y,0)$. The states $\ket{\psi_i}$ thus contain equal
contributions of the spin-split eigenstates
$\ketk{\psi^\mathrm{SIA}_\pm (k_x)}$. Assuming that the electrons in
the source contact are polarized in $+z$ direction we get
(neglecting the envelope functions $\xi_{\kk_\|} (z)$ which are
unimportant in the present discussion)
\begin{subequations}
\begin{equation}
  \label{eq:datta_inj}
  \ket{\psi_i (x=0)} =
\left|\begin{array}{c} 1 \\ 0 \end{array}\right\rangle
 = \frac{1}{2} \left(
  \left|\begin{array}{c}
   1 \\ -i
  \end{array}\right\rangle +
  \left|\begin{array}{c}
   1 \\ i
  \end{array}\right\rangle \right).
\end{equation}
The basis states on the right hand side of Eq.\ (\ref{eq:datta_inj})
propagate with wave vectors $k_x \mp \Delta k/2$ as depicted in
Fig.~\ref{fig:rashba_edisp_pmdk}
\begin{equation}
\label{eq:datta_prec}
  \arraycolsep 0.2ex
  \begin{array}{rl} \displaystyle
  \ket{\psi_i (x)} = \frac{1}{2} \bigg( &
  \exp\left[i\left(k_x - \Delta k/2 \right)x \right]
  \left|\begin{array}{c}
   1 \\ -i
  \end{array}\right\rangle \\[3ex] &
  + \exp\left[i\left(k_x + \Delta k/2 \right)x \right]
  \left|\begin{array}{c}
   1 \\ i
  \end{array}\right\rangle \bigg).
    \end{array}
\end{equation}    
\end{subequations}
Due to the different phase velocities of the basis states in Eq.\
(\ref{eq:datta_prec}) we thus get
\begin{equation}
  \label{eq:datta_exp}
  \expect{\vek{S} (x)} = \left(
    \begin{array}{c}
      \sin (- \Delta k \, x) \\
      0  \\
      \cos (\Delta k \, x)
    \end{array}\right) .
\end{equation}
This equation can be visualized by saying that the spin vector
$\expect{\vek{S}}$ of the state $\ket{\psi_i (x)}$ precesses around
the effective field $\vekc{B} (k_x) = (0,\mathcal{B}_y,0)$ (see
Fig.~\ref{fig:datrans}). Note, however, that conventional spin
precession \cite{sak94} takes place as a function of time~$t$
whereas in Eq.\ (\ref{eq:datta_exp}) the spin precesses as a
function of position $x$.

If finally the drain contact at $x=L$ is ferromagnetic, too, the
electrons can exit the semiconducting channel only if the spin
orientation $\expect{\vek{S} (x=L)}$ of the electrons matches the
polarization $\vek{P}_\mathrm{D}$ of the drain contact,
\begin{equation}
  \label{eq:D_chi}
  \cos \chi =
  \vek{P}_\mathrm{D} \cdot \expect{\vek{S} (x=L)},
\end{equation}
where $\chi$ denotes the angle between $\vek{P}_\mathrm{D}$ and
$\expect{\vek{S} (x=L)}$. A large positive value of $\cos \chi$
indicates that the electrons can easily exit the semiconducting
channel whereas a large negative value indicates that the
spin-polarized current is suppressed. Assuming that
$\vek{P}_\mathrm{S,D} \parallel [001]$ we obtain from Eq.\ 
(\ref{eq:datta_exp})
\begin{equation}
  \label{eq:tran_olap}
  \cos \chi = \cos (\Delta k \, L).
\end{equation}
A tunable device is achieved if the wave vector difference $\Delta
k$ is varied by changing the Rashba coefficient $\alpha$, see Eq.\ 
(\ref{eq:rashba_dk}) and Fig.~\ref{fig:datrans}.

In the above qualitative discussion we have ignored details such as
the resistance mismatch at the interfaces \cite{sch00,ras00,fer01}
which are important for the practical realization of such a device.
But these aspects do not affect the spin precession inside the
semiconducting channel which is the subject of the present
investigation.

%%%%%%%%%%%%%%%%%%%%%%%%%%%%%%%%%%%%%%%%%%%%%%%%%%%%%%%%%%%%%%%%%%
\subsection{Precession in the presence of BIA and SIA}

In the preceeding subsection we have assumed that only the Rashba
term (\ref{rashba}) contributes to spin splitting. Here the
effective magnetic field $\vekc{B}(\kk_\|)$ that characterizes the
spin orientation of the eigenstates (\ref{eq:rashba_eigen}) is
always perpendicular to the direction $\kk_\|$ of propagation in the
spin transistor. In general, we have both SIA and BIA spin splitting
so that the effective field $\vekc{B}(\kk_\|)$ is a more complicated
function of $\kk_\|$, see Fig.~\ref{fig:beff_gainas}. An arbitrarily
oriented effective field $\vekc{B}(\kk_\|)$ can be characterized by
polar angles $\theta$ and $\phi$, i.e., $\vekc{B} = \Delta k \,
[\sin \theta \cos\phi, \sin\theta \sin\phi, \cos\theta]$. The
corresponding orthonormal eigenstates are
\begin{subequations}
\label{eq:sp_eigen_gen}
\begin{eqnarray}
  \ketk{\uparrow} & = & \left(\begin{array}{c}
      e^{-i\phi/2} \cos (\theta/2) \\ e^{i\phi/2} \sin (\theta/2)
  \end{array}\right) 
  \\ \rule{0pt}{4ex}
  \ketk{\downarrow} & = & \left(\begin{array}{c}
      - e^{-i\phi/2} \sin (\theta/2) \\ e^{i\phi/2} \cos (\theta/2)
  \end{array}\right).
\end{eqnarray}  
\end{subequations}
For any values of the angles $\theta$ and $\phi$, the spin states
(\ref{eq:sp_eigen_gen}) represent a basis of the spin $1/2$ space.
Similar to Eq.\ (\ref{eq:datta_inj}) we can thus expand the wave
function $\ket{\psi_i}$ of the spin-polarized electrons in the
ferromagnetic source contact in terms of the basis states
(\ref{eq:sp_eigen_gen})
\begin{subequations}
\label{eq:sp_gen}
\begin{equation}
  \label{eq:sp_expand}
    \ket{\psi_i (x=0)} =
    \cos u \ketk{\uparrow} + \sin u \, e^{iv} \ketk{\downarrow}
\end{equation}
with angles $u$ and $v$. Thus we get for the precessing electrons
inside the channel
\begin{equation}
\label{eq:sp_prec}
  \arraycolsep 0.2ex
  \begin{array}{rl}
  \ket{\psi_i (x)} = &
    \exp\left[i\left(k_x - \Delta k/2 \right)x \right]
  \, \cos u \, \ketk{\uparrow} \\[2ex] &
  + \exp\left[i\left(k_x + \Delta k/2 \right)x \right]
  \, \sin u \, e^{iv} \, \ketk{\downarrow}.
\end{array}
\end{equation}    
\end{subequations}
Then the overlap of the spin vector $\expect{\vek{S} (x)}$ with the
field $\vekc{B}$ is given by
\begin{equation}
  \label{eq:ovlp}
  \vekc{B} \cdot \expect{\vek{S} (x)} = \Delta k \, \cos (2u)
\end{equation}
independent of the position $x$ inside the channel and independent
of the phase $e^{iv}$. This equation shows that in generalization of
Eq.\ (\ref{eq:datta_exp}) the spin is precessing on a cone around
the effective field $\vekc{B}$ where the cone angle is $2u$. The
precession amplitude $\Delta k \, \cos (2u)$ is the largest when
$u=\pi/4$ so that in Eq.\ (\ref{eq:sp_gen}) we have equal
contributions of the spin-split states $\ketk{\uparrow}$ and
$\ketk{\downarrow}$. This corresponds to the situation that the spin
polarization $\vek{P}_\mathrm{S} = \expect{\vek{S}(x=0)}$ in the
ferromagnetic source contact is perpendicular to $\vekc{B}
(\kk_\|)$. Spin precession is suppressed for $u=0$ and $u=\pi/2$
when the spin polarization $\vek{P}_\mathrm{S}$ in the ferromagnetic
source contact is parallel to $\vekc{B} (\kk_\|)$ so that only one
spin state (\ref{eq:sp_eigen_gen}) contributes in Eq.\ 
(\ref{eq:sp_gen}).

We have seen in Fig.~\ref{fig:beff_gainas} that for a fixed wave
vector $\kk_\|$ the orientation of $\vekc{B} (\kk_\|)$ can change
when the Rashba spin-orbit interaction is tuned by means of an
external gate. It follows that the basic operating principle of the
Datta spin transistor remains valid for the more general eigenstates
(\ref{eq:sp_eigen_gen}) provided the polarization
$\vek{P}_\mathrm{S}$ of the ferromagnetic source contact is
orthogonal to $\vekc{B} (\kk_\|)$ for all values of the external
``knob'' that is used to tune the spin-orbit interaction. If the condition
$\vek{P}_\mathrm{S} \perp \vekc{B} (\kk_\|)$ is not strictly
fulfilled the tunability of the spin transistor is reduced.
We note that these conclusions are valid also for the more general
eigenstates (\ref{eq:envelope_qw}).

%%%%%%%%%%%%%%%%%%%%%%%%%%%%%%%%%%%%%%%%%%%%%%%%%%%%%%%%%%%%%%%%%%
\subsection{Numerical Results}

We present next numerically calculated results for the spin
precession in a spin transistor obtained by means of an $8\times 8$
Kane Hamiltonian \cite{tre79,win93a} that takes fully into account both BIA
and SIA. According to Fig.~\ref{fig:beff_gainas} the effective
fields $\vekc{B} (\kk_\|)$ due to BIA and SIA in a [001]-grown QW
are always parallel to each other for $\kk_\| \parallel
[1\overline{1}0]$ and $\kk_\| \parallel [110]$. On the other hand,
for $\kk_\| \parallel [100]$ the fields are perpendicular to each
other so that we want to focus on these two extreme cases. We will
again consider a 100~{\AA} wide Ga$_{0.47}$In$_{0.53}$As-InP QW, and
we assume that the distance between source and drain contact is
$L=5$~$\mu$m. For ease of notation we will use a suitably rotated
coordinate system (Fig.~\ref{fig:datrans}) such that the electrons
always propagate in $x$ direction, i.e., $\kk_\| = (k_x, 0,0)$. We
assume that the Rashba spin-orbit coupling is tuned by applying an electric
field $\Ee$ perpendicular to the plane of the quasi 2D system
(Fig.~\ref{fig:datrans}).

\begin{figure}[tbp]
  \includegraphics[height=0.38\textheight]{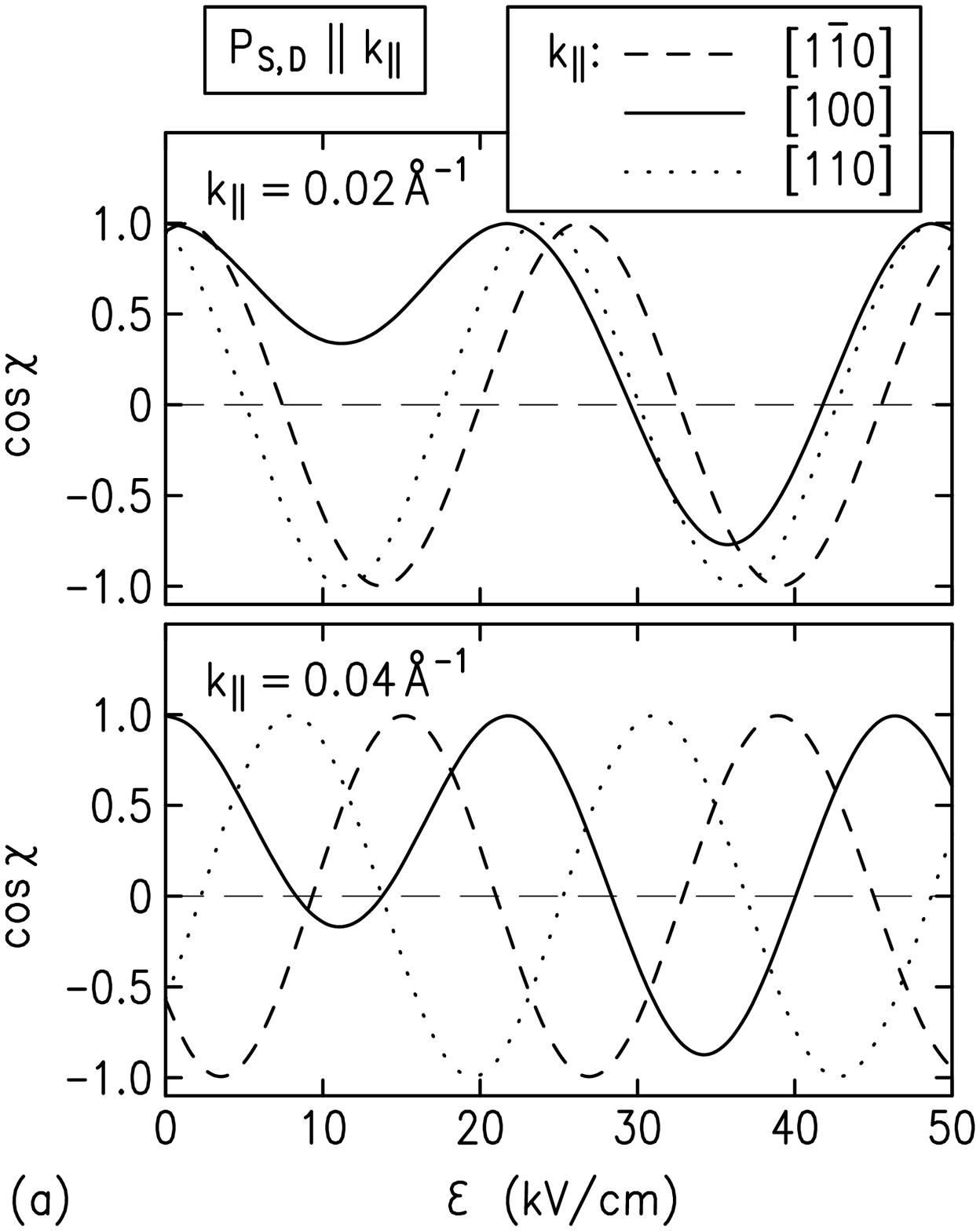}

  \vspace{2ex}
  \includegraphics[height=0.38\textheight]{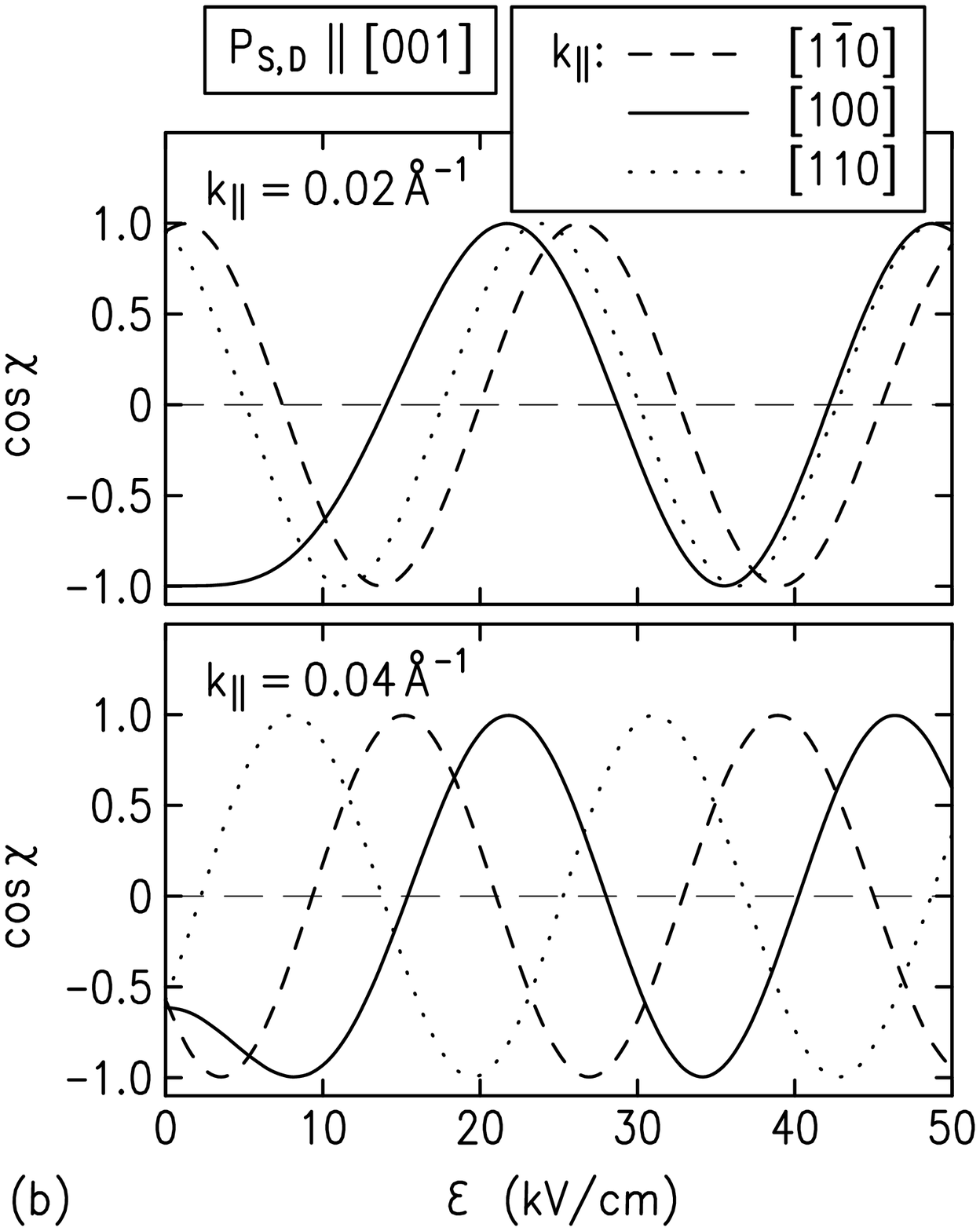}

  \caption[]{\label{fig:D_chi} Overlap $\cos\chi$ between the spin vector
  $\expect{\vek{S} (x=L)}$ and the polarization $\vek{P}_\mathrm{D}$
  of the drain contact as a function of electric field $\mathcal{E}$
  in a 100~{\AA} wide Ga$_{0.47}$In$_{0.53}$As-InP QW with a channel
  length of $L=5$~$\mu$m. In (a) we assume $\vek{P}_\mathrm{S,D}
  \parallel \kk_\|$ whereas in (b) we assume $\vek{P}_\mathrm{S,D}
  \parallel [001]$.  Different line styles correspond to different
  crystallographic directions of $\kk_\|$ as indicated. The
  calculations are based on an $8\times 8$ Kane Hamiltonian
  ($\Gamma_6^c$, $\Gamma_8^v$, and $\Gamma_7^v$) including
  off-diagonal remote band contributions of second order in~$\kk$
  (Ref.~\onlinecite{tre79,win93a}).}
\end{figure}

In Fig.~\ref{fig:D_chi} we show the overlap $\cos \chi$ between the
spin vector $\expect{\vek{S} (x=L)}$ and the polarization
$\vek{P}_\mathrm{D}$ of the drain contact as a function of electric
field $\mathcal{E}$.
We consider different polarization states $\vek{P}_\mathrm{S}$ of
the source contact and it is assumed that $\vek{P}_\mathrm{S}
\parallel \vek{P}_\mathrm{D}$. The results in Fig.~\ref{fig:D_chi}
can readily be understood by means of Fig.~\ref{fig:beff_gainas}.
(i) If $\kk_\| \parallel [1\overline{1}0]$ or $\kk_\| \parallel
[110]$ BIA is of little importance because $\vekc{B}_\mathrm{BIA}
\parallel \vekc{B}_\mathrm{SIA}$. Consistent with Eq.\ 
(\ref{eq:tran_olap}) we thus get a sinusoidal dependence of $\cos
\chi$ on $\Ee$ with the same angle $\chi$ for $\vek{P}_\mathrm{S,D}
\parallel \kk_\|$ and $\vek{P}_\mathrm{S,D} \parallel [001]$, see
Figs.~\ref{fig:D_chi}(a) and (b). We note that for fixed magnitudes
of $\kk_\|$ and $\Ee$ the angle $\chi$ can be adjusted by changing
the length $L$ of the channel. In the present work $L$ has not been
optimized. Note that the smaller is the length $L$ the larger must
be the modulation of $\Ee$ for switching the device. (ii) For
$\vek{P}_\mathrm{S,D} \parallel \kk_\| \parallel [100]$ and $\Ee =
0$ the spin precession is suppressed because $\vek{P}_\mathrm{S,D}
\parallel \vekc{B}$. In this case we have $\cos\chi = 1$ independent
of the channel length $L$. For $\Ee >0$ the spin states start to
precess. Here spin precession and $\cos \chi$ are more complicated
functions of $\Ee$ because $\Ee$ changes both the magnitude and
orientation of $\vekc{B}$. (iii) For a QW grown in the high-symmetry
crystallographic direction $[001]$ the overlap $\cos\chi$ is
symmetric with respect to $\Ee > 0$ and $\Ee < 0$. In the latter
case the roles of $\kk_\| \parallel [1\overline{1}0]$ and $\kk_\|
\parallel [110]$ are reversed.

\begin{figure}[tbp]
  \includegraphics[width=0.8\columnwidth]{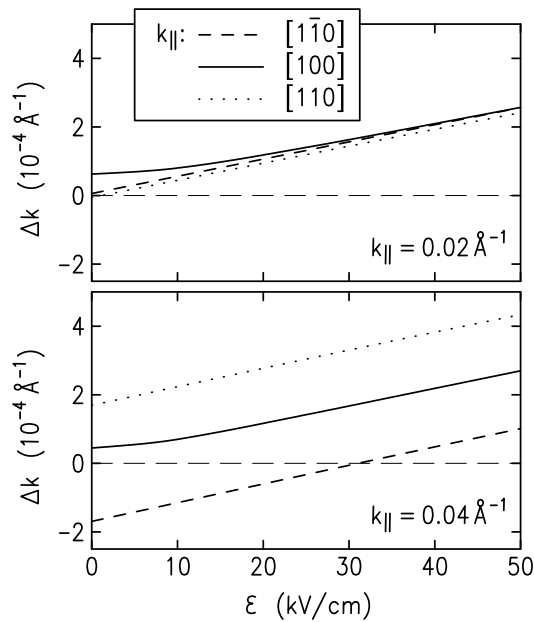}
  \caption[]{\label{fig:Dkvek} Spin splitting $\Delta k$ as a
  function of electric field $\mathcal{E}$ in a 100~{\AA} wide
  Ga$_{0.47}$In$_{0.53}$As-InP QW. We consider different magnitudes
  and different crystallographic directions of $\kk_\|$ as indicated
  in the figure.}
\end{figure}

It is interesting to compare Fig.~\ref{fig:D_chi} with the magnitude
of spin splitting $\Delta k$ as a function of electric field
$\mathcal{E}$ (Fig.~\ref{fig:Dkvek}).
We see that $\Delta k$ depends rather sensitively on both the
magnitude and orientation of the wave vector $\kk_\|$. Nevertheless,
we obtain in Fig.~\ref{fig:D_chi} the same modulation of the overlap
$\cos \chi$ as a function of $\Ee$ for $\kk_\| \parallel
[1\overline{1}0]$ and $\kk_\| \parallel [110]$, independent of the
magnitude of $\kk_\|$ [apart from a constant phase shift $\chi_0
(\kk_\|)$]. This is due to the fact that the relevant quantity for
the spin transistor is not the absolute value $\Delta k$ of the spin
splitting, but the {\em variation} $\partial (\Delta k) / \partial
\mathcal{E}$. We see in Fig.~\ref{fig:Dkvek} that the latter
quantity depends much more weakly on the magnitude and orientation
of $\kk_\|$. Furthermore, it is advantageous that the orientation of
$\vekc{B}(\kk_\|)$ is independent of the magnitude of the tunable
part of the spin-orbit interaction. It can be seen in
Fig.~\ref{fig:beff_gainas} that this condition is fulfilled for
$\kk_\| \parallel [1\overline{1}0]$ and $\kk_\| \parallel [110]$ but
not for $\kk_\| \parallel [100]$. Therefore the modulation of
$\cos\chi$ as a function of $\Ee$ is more pronounced in the former
case than for $\kk_\| \parallel [100]$, even though in all cases the
spin splitting $\Delta k$ shows roughly the same field dependence
$\partial (\Delta k) / \partial \mathcal{E}$.

%%%%%%%%%%%%%%%%%%%%%%%%%%%%%%%%%%%%%%%%%%%%%%%%%%%%%%%%%%%%%%%%%%
\subsection{Spin Precession and Spin Relaxation}

For the Datta spin transistor it is advantageous to have a small
spin relaxation in the semiconducting channel because spin
relaxation is competing with the controlled spin precession in the
channel. Typically, the dominant mechanism for spin relaxation in 2D
electron systems is the one proposed by D'yakonov and Perel' (DP).
\cite{dya72,dya86} It can be viewed as a spin precession in the
effective field $\vekc{B}$ that is randomized because $\vekc{B}$
changes when momentum scattering changes the wave vector $\kk_\|$ of
the electrons. DP spin relaxation can therefore be suppressed if
(apart from a sign of $\vekc{B}$) the orientation of $\vekc{B}$ is
independent of the wave vector $\kk_\|$ and the spins of the
propagating electrons are oriented parallel to $\vekc{B}$. Such a
situation can be realized in a symmetric QW grown in the
crystallographic direction $[110]$ where $\vekc{B}$ is perpendicular
to the plane of the QW, \cite{dya86,ohn99} see
Fig.~\ref{fig:beff_gainas}(e). Similarly, in a QW grown in the
crystallographic direction $[001]$ with $|\alpha| = |\eta|$ we have
in first order of $\kk_\|$ that $\vekc{B} \parallel [110]$ (or
$\vekc{B} \parallel [1\overline{1}0]$ depending on the sign of
$\alpha$ and $\eta$). \cite{sch03} In both cases spin relaxation is
suppressed only for a particular value of the Rashba spin-orbit
coupling (i.e., a particular value of the field $\Ee$). For the spin
transistor it is preferable to have a regime of electric fields
$\Ee$ with suppressed spin relaxation so that we can switch between
$\cos \chi = 1$ and $\cos \chi = -1$.

Recently, an alternative spin transistor has been proposed
\cite{sch03} that is less sensitive to spin relaxation. It uses the
fact that not only DP spin relaxation can be suppressed if (apart
from a sign of $\vekc{B}$) the orientation of the effective field
$\vekc{B}$ is the same for all wave vectors $\kk_\|$; but obviously
spin precession is then suppressed, too. Therefore, if $\vekc{B}
\parallel \vek{P}_\mathrm{S,D}$ electrons travel unperturbed through
the device which corresponds to the ``on'' state. In a detuned
system, on the other hand, $\vekc{B}$ varies as a function of
$\kk_\|$ which implies that, in general, $\vekc{B} \nparallel
\vek{P}_\mathrm{S,D}$. Therefore, spin precession and/or DP spin
relaxation reorient the spins in the channel. The spin vector
$\expectk{\vek{S}(x=L)}$ thus no longer matches the polarization
$\vek{P}_\mathrm{D}$ of the drain contact so that the current
through the device diminishes.

Such a spin transistor can be built using a QW grown in the
crystallographic direction [110]. Here it follows from
Fig.~\ref{fig:beff_gainas}(e) that if the QW is symmetric then
DP spin relaxation is suppressed because $\vekc{B}$ is
perpendicular to the plane of the QW for all in-plane wave vectors
$\kk_\|$. If $\vek{P}_\mathrm{S,D} \parallel \vekc{B}$ electrons
thus travel unperturbed through the device. If the QW is made
asymmetric by applying an electric field $\Ee$ perpendicular to the
plane of the well, the current diminishes because of the onset of
spin precession and/or DP spin relaxation.

Alternatively, we can use a QW grown in the crystallographic
direction [001] (Ref.\ \onlinecite{sch03}). Here we can achieve in
linear order of $\kk_\|$ that $\vekc{B}$ is independent of $\kk_\|$
if $|\alpha| = |\eta|$. This situation is approximately shown by the
innermost contour in Fig.~\ref{fig:beff_gainas}(d). Note, however,
that higher orders in spin splitting [in particular the cubic term
in Eq.\ (\ref{eq:bia_sia})] do not comply with the requirement that
in the on state of the device the orientation of $\vekc{B}$ should
be independent of $\kk_\|$. Furthermore, we see in
Fig.~\ref{fig:beff_gainas}(d) that only the orientation but not the
magnitude of $\vekc{B}$ is independent of $\kk_\|$. For electrons
with $\kk_\| \parallel [110]$ we have actually $\mathcal{B}=0$
whereas $\mathcal{B}$ is the largest for $\kk_\| \parallel
[1\overline{1}0]$. In the former case (i.e., for
$\vek{P}_\mathrm{S,D} \parallel \kk_\| \parallel [110]$) changing
$\Ee$ diminishes the current through the device because we have then
$\vek{P}_\mathrm{S,D} \perp \vekc{B}$ so that injected electrons
precess around $\vekc{B}$. The electrons do not precess in the
latter case because we have $\vek{P}_\mathrm{S,D} \parallel
\vekc{B}$ independent of $\Ee$. DP spin relaxation is highly
anisotropic, too. Here the situation is actually reversed: We have
large spin relaxation rates for those directions of
$\expect{\vek{S}}$ for which we have a large $\vekc{B} (\kk_\|)$
(Refs.~\onlinecite{ave99,kai03a}). Therefore, spin relaxation
supports the switching of the device most effectively if $\kk_\|
\parallel [1\overline{1}0]$. We remark that an all-inclusive
investigation of this question should explicitely evaluate spin
relaxation as a function of $\kk_\|$ and $\expectk{\vek{S}}$.

%%%%%%%%%%%%%%%%%%%%%%%%%%%%%%%%%%%%%%%%%%%%%%%%%%%%%%%%%%%%%%%%%%
\section{Conclusions}

In general, the total $B=0$ spin splitting in inversion asymmetric
2D systems is determined by an interplay of spin splitting due to
BIA, which is always present in systems with a zinc blende
structure, and the tunable spin splitting due to SIA. These spin
splittings can be characterized by effective magnetic fields
$\vekc{B}(\kk_\|)$ that vary as a function of in-plane wave vector
$\kk_\|$. The functional form of $\vekc{B}_\mathrm{SIA} (\kk_\|)$
due to SIA is independent of the crystallographic direction in which
a QW has been grown. Due to the axial symmetry of the Rashba term
the field $\vekc{B}_\mathrm{SIA} (\kk_\|)$ is always perpendicular
to $\kk_\|$ in the plane of the QW. Furthermore, it is only weakly
dependent on the magnitude of $\kk_\|$. On the other hand, the field
$\vekc{B}_\mathrm{BIA} (\kk_\|)$ due to BIA depends sensitively both
on the magnitude and orientation of $\kk_\|$ as well as on the
crystallographic direction in which the QW was grown. For QW's grown
in the direction [001] the field $\vekc{B}_\mathrm{BIA} (\kk_\|)$ is
always in the plane of the QW whereas for QW's grown in the
direction [110] it is pointing perpendicular to the plain of the QW.
For other growth directions the field $\vekc{B}_\mathrm{BIA}
(\kk_\|)$ has both in-plane and out-of-plane components.

Electrons injected into a 2D semiconducting channel propagate with a
certain in-plane wave vector $\kk_\|$. If these electrons are
spin-polarized such that the spinor $\ket{\psi_i}$ of the electrons
is not a spin eigenstate of the system, the spin of the propagating
electrons precesses in the effective field $\vekc{B}(\kk_\|)$. The
precession is the largest if the spin orientation
$\expect{\vek{S}}$ of the electrons is perpendicular to the
effective field $\vekc{B}(\kk_\|)$. In a QW grown in the
crystallographic direction [001] it is thus advantageous that the
electrons are injected in the in-plane directions $[1\overline{1}0]$
or $[110]$ because here the field $\vekc{B}(\kk_\|)$ is always
perpendicular to the direction of propagation. For the direction
[100], on the other hand, the fields due to BIA and SIA are
perpendicular to each other so that the orientation of the total
field $\vekc{B}(\kk_\|)$ depends on the magnitude of BIA and SIA
spin splitting.

%%%%%%%%%%%%%%%%%%%%%%%%%%%%%%%%%%%%%%%%%%%%%%%%%%%%%%%%%%%%%%%%%%
\begin{acknowledgments}
  The author would like to thank U.\ Merkt and G.\ Meier for
  stimulating discussions and suggestions.
\end{acknowledgments}

%%%%%%%%%%%%%%%%%%%%%%%%%%%%%%%%%%%%%%%%%%%%%%%%%%%%%%%%%%%%%%%%%%


\begin{thebibliography}{44}
\bibitem[{ire()}]{irep}
\bibinfo{note}{From a group theoretical point of view this can be traced back
  to the fact that $\ketk{\psi_+ (\kk_\|)}$ and $\ketk{\psi_- (\kk_\|)}$
  transform according to different irreducible representations of the group of
  the wave vector $\kk_\|$.}
\bibitem[{c2v()}]{c2v}
\bibinfo{note}{For a symmetric QW grown in the crystallographic direction [110]
  the symmetry axis of the point group $C_{2v}$ is parallel to the axis $[001]$
  in the plane of the QW.}
\end{thebibliography}
\end{document}